\documentclass[11pt,a4paper]{article}
\usepackage[T1]{fontenc}
\usepackage{lmodern}
\usepackage[british]{babel}
\usepackage{amsmath,amssymb,bm}
\usepackage{graphicx}
\usepackage{xcolor}
\usepackage{geometry}
\usepackage{microtype}
\usepackage{booktabs}
\usepackage[section]{placeins}
\usepackage{hyperref}
\hypersetup{colorlinks=true,linkcolor=black,citecolor=black,urlcolor=blue}
\numberwithin{equation}{section}

\title{A possible signature of neutrino spin precession in the J-PARC muon $g-2$ experiment}
\author{S. Carneiro$^{1}$\footnote{saulocarneiro@on.br\\ \indent ORCID: 0000-0001-7098-383X}\; and F. C. Sobrinho$^{2}$\\[0.5em]
\small $^{1}$Observat\'orio Nacional, 20921-400 Rio de Janeiro, RJ, Brazil\\
\small $^{2}$Instituto de F\'isica, Universidade de S\~ao Paulo, 05508-090 S\~ao Paulo, SP, Brazil}
\begin{document}
\maketitle

\begin{abstract}
We examine a possible signature of the neutrino angular-momentum contribution to the muon magnetic anomaly measurement in the J-PARC E34 experiment. In previous papers we argued that the angular momentum carried by the unobserved neutrino pair may leave an uncertainty in the anomaly inferred from the time modulation of detected positrons. At Fermilab the highly relativistic muons make the forward-emission approximation suitable for this estimate. The lower muon momentum at J-PARC requires the decay kinematics to be treated explicitly. We integrate the longitudinal momentum of the neutrino pair using the positron energy distribution and asymmetry derived from the theoretical weak-decay spectrum, and the same event-counting weight $N(E)A(E)$ is employed for both experiments. For positron energies between $1.5$ and $2.9$ GeV at Fermilab and between $200$ and $275$ MeV at J-PARC, the proposed correction is smaller at J-PARC by $0.415\pm0.176$ ppm, when the most optimistic E34 precision is assumed. The expected difference amounts to $2.36$ standard deviations if the future measurement is centred on this value. The calculation thus identifies an experimental comparison which may test the proposed effect independently of the hadronic contribution to the Standard Model prediction. We also examine the possibility of testing the effect by comparing measurements at two different muon momenta. With a fixed positron energy threshold of 200 MeV, the maximum sensitivity is obtained at a muon beam energy of 245 MeV. The predicted difference relative to the nominal J-PARC configuration has significance of $4.88\sigma$ for the most optimistic experimental sensitivity.

\end{abstract}

\section{Introduction}
The magnetic anomaly of the muon remains one of the most precise tests of the Standard Model. Its measurement at Brookhaven and Fermilab is based on the time modulation of the positrons emitted by polarised muons precessing in a magnetic storage ring \cite{bennett,aguillard2025,fnalreport}. The last Fermilab result has a relative uncertainty of $127$ parts per billion \cite{aguillard2025}, which has decreased considerably since the first Brookhaven measurements. At the same time, the theoretical situation has also changed since then.

The traditional dispersive determination of the leading hadronic vacuum polarisation uses measured $e^+e^-\to\mathrm{hadrons}$ cross sections \cite{aoyama2020,davier2020}. When these cross sections give a relatively low hadronic contribution, the resulting magnetic anomaly lies below the experimental value. The dispersive estimates themselves are affected by the differences between the $\pi^+\pi^-$ cross sections reported by BaBar, KLOE and CMD-3 experiments \cite{babar,kloe,cmd3}. The present experimental precision therefore makes the disagreement among the hadronic inputs particularly relevant.

The hadronic contribution is in fact the main source of theoretical uncertainty in the magnetic anomaly $a_\mu$. While the dispersion relation connects the hadronic vacuum polarisation to experimentally determined cross sections, lattice QCD calculates the same contribution from correlation functions in a non-perturbative formulation of the theory \cite{aoyama2020,borsanyi2021,aliberti2025}.
These approaches have not always led to compatible results. The first high-precision lattice determination by Borsanyi \emph{et al.} was closer to the measured anomaly than the earlier dispersive evaluations \cite{borsanyi2021}. The updated 2025 White Paper, which incorporates lattice results in its recommended evaluation, quotes a value that
agrees with the experimental average within one standard deviation \cite{aliberti2025}. This agreement, however, does not settle the discrepancies among the dispersive inputs. The status of the calculation is reviewed in Ref.~\cite{aliberti2025}.

In two previous papers we considered another source of uncertainty which could affect the inference of $a_\mu$ from the observed positron signal \cite{carneiro2023,carneiro2026}. A subsequent work examined its predicted energy dependence using public scans from the Fermilab collaboration \cite{carneiro2026pre}. Here we address a different experimental test. The J-PARC E34 experiment will store muons with a momentum far below the magic momentum employed at Fermilab \cite{abe2019,otani2022,e34status}, and the change in decay kinematics gives a direct way of comparing the proposed correction at two different muon energies.

Our proposal concerns a possible contribution to the experimentally inferred anomaly, not to theoretical estimations. Its comparison with the J-PARC measurement can therefore be made without selecting a particular hadronic evaluation. As we shall see, the use of a second muon momentum is particularly valuable for this purpose.

\section{The neutrino trace and the anomaly correction}
The measured positron distribution originates in the weak decay
\begin{equation}
 \mu^+\longrightarrow e^+ +\nu_e+\bar\nu_\mu.
\end{equation}
The final state contains correlations between the momenta and spins of the three decay products. The experimental observable, however, is reconstructed from an ensemble of positrons detected at different times and energies. The two neutrinos remain unobserved.

In our previous papers we questioned the usual inference obtained by tracing over the neutrino helicities at the decay vertex \cite{carneiro2026,carneiro2026pre}. The trace relevant to the measurement is defined on the asymptotic state associated with the detected positron ensemble, and it may in principle be affected by the angular momentum exchanged with the external field during the flight of the unobserved particles. In the relativistic decay, a trace defined at the vertex cannot simply be identified with a trace of the detector-defined out state by evolving backwards a fixed helicity basis \cite{Landau 4}. The assumptions underlying that identification have to be established for the observed state, including its momentum correlations.
We retain here the angular-momentum estimate derived in Refs.~\cite{carneiro2026,carneiro2026pre}. The present paper is concerned with its kinematic consequence for an experiment using less relativistic muons.

Consider a neutrino whose spin $\mathbf{s}$ is initially aligned with its momentum and whose trajectory, along the $x$-direction, lies in the plane perpendicular to a uniform magnetic field pointing to the $z$-direction. The $y$-component of its laboratory-frame angular momentum follows from the Lorentz transformation of its tensor components \cite{carneiro2026,landau1975},
\begin{equation}
 L_y=\gamma_\nu\bigl(s_y-v_\nu L^{\prime03}\bigr),
 \qquad L^{\prime03}=t'p'_z-E'z',
 \label{eq:ly}
\end{equation}
where the prime refers to the neutrino rest frame.
For free motion $L^{\prime03}$ is constant. The neutrino angular momentum then acquires a variation 
\begin{equation}
 \delta L_y= \gamma_\nu \delta s_y = -\gamma_\nu\mu_\nu B\,dt,
 \label{eq:dl}
\end{equation}
where $\mu_\nu$ is its magnetic moment.
In the forward approximation used in our earlier treatment the two neutrinos then contribute coherently to an angular-momentum uncertainty after tracing. Balancing their contribution with the inferred muon precession gives 
\begin{equation}
\gamma_\mu\delta\mu_\mu=2\gamma_\nu\mu_\nu,
 \label{eq:balance}
\end{equation}
where $\delta \mu_\mu$ is the resulting correction in the muon magnetic moment.

For a Dirac massive neutrino, the minimal extension of the electroweak model gives \cite{giunti2015}
\begin{equation}
 \mu_\nu=\frac{3eG_Fm_\nu}{8\sqrt{2}\pi^2},
\end{equation}
and we obtain the correction introduced in Ref.~\cite{carneiro2026} for the muon magnetic anomaly,
\begin{equation}
 \delta a_\mu=\frac{3m_\mu^2G_F}{4\sqrt{2}\pi^2}f
 \simeq 7.0\times10^{-9} f.
 \label{eq:delta}
\end{equation}
Here $f$ measures the fraction of the muon energy carried by the neutrino pair.

Our previous comparison with data-driven calculations used the asymmetry-weighted correction
\begin{equation}
 \overline{\delta a}_\mu^{\,A}
 =\frac{\int dE\,A(E)\delta a_\mu(E)}{\int dE\,A(E)}.
 \label{eq:oldaverage}
\end{equation}
In the interval $1.5<E<2.9$ GeV, the theoretical decay spectrum used here gives $\langle f\rangle_A=0.21523$ and a correction $1.507\times10^{-9}$, or about $1.29$ ppm. Subtracting this estimate from the final Fermilab value gives a corrected central value $a_\mu\times10^{10}-11659000=191.98$. For reference, the BaBar-based dispersive result lies at $189.10\pm4.80$ on the same scale. These values agree within the quoted uncertainty of the latter \cite{babar,carneiro2026}.

It is worth distinguishing the asymmetry weight in Eq.~(\ref{eq:oldaverage}) from the event-by-event asymmetry weighting used in Fermilab's A method. The latter has an effective spectral weight $N(E)A(E)^2$. With the present ideal decay spectrum and the same energy interval, it gives $\langle f\rangle_{NA^2}=0.21940$, only about $1.9\%$ above the result of Eq.~(\ref{eq:oldaverage}). The proximity of the two estimates shows that the conclusion of our earlier comparison is little affected by this distinction. The precise experimental acceptance was not included in either calculation.

Equation~(\ref{eq:delta}) predicts a dependence on the positron energy because the neutrino pair carries the remaining energy and momentum of the decay. In Ref.~\cite{carneiro2026pre} we compared this dependence with published energy-binned $R(E)$ scans for Fermilab runs 1 to 6. The points were digitised from publicly available theses and the original bin-to-bin covariance matrices were not available.
The global analysis used
\begin{equation}
\omega_a=\omega_{a0}\left(1+c\frac{\delta a_\mu(E)}{a_\mu}\right),
\end{equation}
and it gave $c=0.527\pm0.450$ with $\chi^2/\nu=1.055$. The fitted coefficient lies between the constant case $c=0$ and the maximal energy-dependent case $c=1$. The difference in the quality of the fits is small and the existing binned results give only weak support for an energy-dependent contribution \cite{carneiro2026pre}. A comparison between Fermilab and J-PARC offers a separate test with muon beams of substantially different energies.

\section{A forecast for J-PARC E34}
The Fermilab experiment stores muons at approximately $p_\mu=3.094$ GeV/$c$. The E34 design employs muons accelerated to $p_\mu=300$ MeV/$c$ \cite{fnalreport,otani2022}. Taking $m_\mu=105.6584$ MeV, the corresponding Lorentz factors and velocities are
\begin{equation}
\begin{aligned}
 \gamma_{\rm FNAL}&=29.30,&\quad\beta_{\rm FNAL}&=0.999417,\\
 \gamma_{\rm E34}&=3.010,&\quad\beta_{\rm E34}&=0.943211.
\end{aligned}
 \label{eq:gamma}
\end{equation}
At Fermilab the emitted high-energy positrons and neutrinos are confined to a narrow forward cone. At J-PARC the lower boost makes the emission appreciably less collinear with the muon momentum. We therefore express the correction through the component of the neutrino-pair momentum along the muon trajectory.

Let $E_\mu$ and $\bm p_\mu$ be the muon energy and momentum in the laboratory. Momentum conservation gives
\begin{equation}
 \bm p_{\nu_e}+\bm p_{\bar\nu_\mu}=\bm p_\mu-\bm p_e.
 \label{eq:momentum}
\end{equation}
Writing $\theta$ for the angle between the positron and muon momenta, the longitudinal fraction is
\begin{equation}
 f_\parallel
 =\frac{(\bm p_{\nu_e}+\bm p_{\bar\nu_\mu})\cdot\widehat{\bm p}_\mu}{E_\mu}
 =\beta_\mu-\frac{p_e}{E_\mu}\cos\theta.
 \label{eq:flab}
\end{equation}
The expression does not require the individual neutrino momenta to be parallel. Their vector sum is fixed by the observed positron momentum.

To perform the average we use $x=2E_e^*/m_\mu$ and $u=\cos\theta^*$, where the star denotes the muon rest frame and $\theta^*$ is measured from the boost direction. Neglecting the positron mass, the boost gives
\begin{equation}
 y\equiv\frac{E_e}{E_\mu}=\frac{x}{2}(1+\beta_\mu u),
 \qquad
 \frac{p_{e\parallel}}{E_\mu}=\frac{x}{2}(\beta_\mu+u).
 \label{eq:boost}
\end{equation}
Consequently,
\begin{equation}
 f_\parallel(x,u)=\beta_\mu-\frac{x}{2}(\beta_\mu+u).
 \label{eq:fxu}
\end{equation}

The energy distribution and analysing asymmetry follow from the weak-decay spectrum, whose early derivations were given by Tiomno and Wheeler and by Michel \cite{tiomno1949,michel1950}. The spin-dependent spectrum is proportional to
\begin{equation}
 \frac{d^2\Gamma}{dx\,du}
 \propto x^2\Big[(3-2x)+P_\mu(2x-1)u\Big],
 \label{eq:michel}
\end{equation}
where $P_\mu$ denotes the longitudinal component of the muon polarisation along its direction of motion. After transforming to the observed energy $y$, define
\begin{equation}
 \begin{split}
 N(y)&\propto\int_{x_-}^{x_+}\frac{2x}{\beta_\mu}(3-2x)\,dx,\\
 N(y)A(y)&\propto\int_{x_-}^{x_+}\frac{2x}{\beta_\mu}
 (2x-1)\,u(x,y)\,dx,
 \end{split}
 \label{eq:NA}
\end{equation}
where
\begin{equation}
 u(x,y)=\frac{2y/x-1}{\beta_\mu},\qquad
 x_- =\frac{2y}{1+\beta_\mu},\qquad
 x_+=\min\!\left(1,\frac{2y}{1-\beta_\mu}\right).
 \label{eq:limits}
\end{equation}

For an unweighted time spectrum, each energy interval contributes to the oscillating signal through the number of positrons $N(E)$ multiplied by its analysing asymmetry $A(E)$. We use the same effective weight $NA$ at Fermilab and J-PARC. This corresponds to a T-type counting prescription with prescribed energy windows. The published Fermilab average was obtained with asymmetry-weighted methods and is used here because it is the most precise available experimental reference \cite{aguillard2025,fnalreport}. This difference in estimator is a limitation of the present forecast.
We use the theoretical $N$ and $A$ of Eq.~(\ref{eq:michel}) in both experiments. Their observed distributions will also depend on detector response, geometric acceptance and reconstruction thresholds. Those effects are not available in sufficient detail to enter the forecast at present.

For completeness, introduce the kernel
\begin{equation}
 K(x,y;\beta)=\frac{2x}{\beta}(2x-1)
 \frac{2y/x-1}{\beta}.
 \label{eq:kernel}
\end{equation}
The longitudinal fraction averaged with the $NA$ weight is then
\begin{equation}
 {\displaystyle
 \langle f_\parallel\rangle_{NA}=
\frac{\int_{y_-}^{y_+}dy\int_{x_-}^{x_+}dx\,
 K(x,y;\beta)\left[\beta-\frac{x}{2}(\beta+u)\right]}
 {\int_{y_-}^{y_+}dy\int_{x_-}^{x_+}dx\,K(x,y;\beta)}.}
 \label{eq:integral}
\end{equation}
The energy limits $y_-$ and $y_+$ are the corresponding laboratory limits divided by $E_\mu$. Equation~(\ref{eq:integral}) includes the angular correlation of the decay and can be evaluated by ordinary numerical quadrature.
In the phase-averaged estimate adopted here, the transverse components of the neutrino-pair momentum cancel upon averaging over the theoretical decay distribution and the collective muon precession. Their contribution to the angular-momentum uncertainty therefore vanishes, and only the longitudinal component is retained in Eq.~(\ref{eq:integral}).

We select the high-sensitivity interval $1.5<E_e<2.9$ GeV at Fermilab and the simulated E34 interval $200<E_e<275$ MeV \cite{carneiro2026pre,otani2022,e34status}. Equation~(\ref{eq:integral}) yields
\begin{equation}
 \langle f_\parallel\rangle_{\rm FNAL}=0.27542,
 \qquad
 \langle f_\parallel\rangle_{\rm E34}=0.20627.
 \label{eq:averages}
\end{equation}
From Eq.~(\ref{eq:delta}), these correspond to
\begin{equation}
 \begin{aligned}
 \overline{\delta a}_\mu^{\rm FNAL}&=19.28\times10^{-10},\\
 \overline{\delta a}_\mu^{\rm E34}&=14.44\times10^{-10},
 \end{aligned}
 \label{eq:corrections}
\end{equation}
so that
\begin{equation}
 {\Delta a_\mu\equiv
 \overline{\delta a}_\mu^{\rm E34}-
 \overline{\delta a}_\mu^{\rm FNAL}
 =-4.84\times10^{-10}=-0.415\ \mathrm{ppm}.}
 \label{eq:diff}
\end{equation}
The coefficient in Eq.~(\ref{eq:delta}) represents a maximal correction as adopted in our earlier estimates. The forecast is conditional on that amplitude and on the theoretical decay distributions.

\begin{figure}[t]
 \centering
 \includegraphics[width=0.93\textwidth]{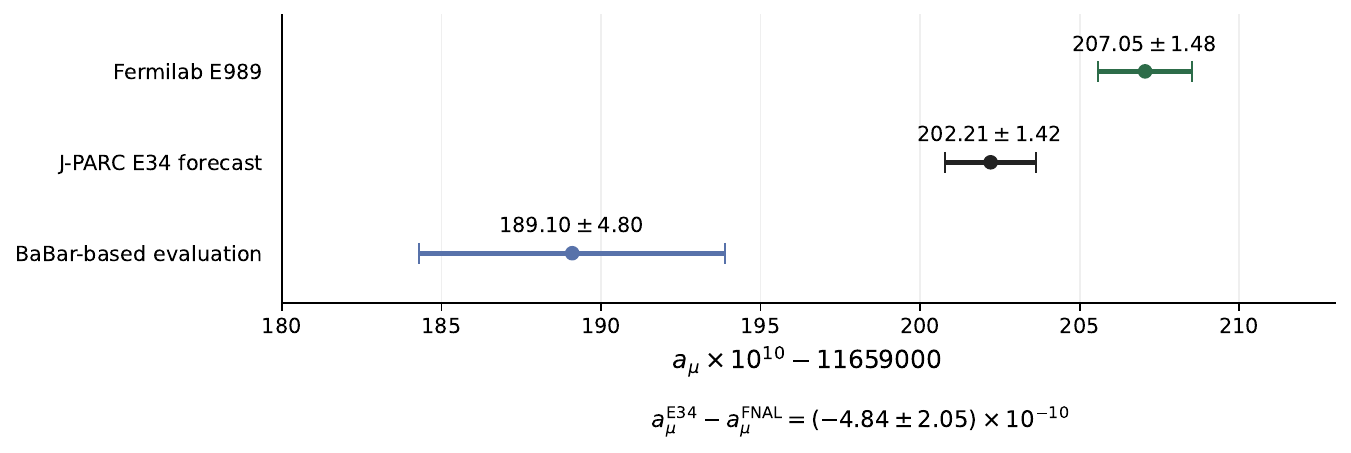}
 \caption{The final Fermilab measurement and our conditional forecast for J-PARC E34. The BaBar-based dispersive value is shown for visual reference only. The horizontal bars denote one standard deviation.}
 \label{fig:forecast}
\end{figure}

The original E34 design envisages a statistical precision of $450$ ppb and a systematic contribution below $70$ ppb{\hypersetup{citecolor=black}~\cite{abe2019}}. An advanced sensitivity near $100$ ppb statistical has also been discussed \cite{otani2022,e34status}. We use the latter as the optimistic scenario. Adding $100$ and $70$ ppb in quadrature gives a projected E34 uncertainty of $0.122$ ppm. Combining it with Fermilab's $0.127$ ppm uncertainty gives
\begin{equation}
 \sigma_{\rm E34-FNAL}
 =\sqrt{0.127^2+0.100^2+0.070^2}
 =0.176\ \mathrm{ppm}.
 \label{eq:error}
\end{equation}
Our final difference is therefore
\begin{equation}
 {\Delta a_\mu=(-4.84\pm2.05)\times10^{-10}
 =(-0.415\pm0.176)\ \mathrm{ppm}.}
 \label{eq:result}
\end{equation}
The separation of the central values corresponds to $2.36$ standard deviations at the optimistic precision. Figure~\ref{fig:forecast} presents the comparison, with the BaBar-based reference making visible the scale of the earlier dispersive estimates.

The predicted difference would be smaller if we used the $NA^2$ weight of Fermilab's A method in both experiments. With the same ideal spectra and windows it would be $-0.192$ ppm. At the initial E34 statistical precision of $450$ ppb, the combined experimental uncertainty would instead be approximately $0.473$ ppm. Neither of these comparisons would provide the sensitivity obtained with the present optimistic $NA$ forecast. A final experimental test should use the measured energy-dependent acceptances and the extraction weights actually employed by each collaboration.

\section{A two-momentum test at J-PARC}
The absence of electrostatic focusing in the E34 storage ring removes the need to operate at the magic muon momentum \cite{otani2022,e34status}. This offers another possible test of the proposed effect. Two measurements made with different muon momenta would share the same experimental method and could be compared without using the Fermilab result as an experimental reference. We consider the existing design momentum of $300$ MeV/$c$ and a second, lower momentum. Changing the momentum would also change the orbit and the detector acceptance. Our calculation describes the sensitivity that might be obtained in a dedicated experiment with suitable injection and detection systems.

We keep the positron threshold fixed at $E_{\min}=200$ MeV and impose $E_{\max}=275$ MeV as an upper selection limit{\hypersetup{citecolor=black}~\cite{abe2019}}. The latter becomes immaterial when the kinematic endpoint lies below $275$ MeV. The same theoretical decay spectrum and $NA$ weight used above are retained. For a trial momentum $p$, the total muon energy and velocity are
\begin{equation}
 E_\mu(p)=\sqrt{p^2+m_\mu^2},\qquad
 \gamma(p)=\frac{E_\mu(p)}{m_\mu},\qquad
 \beta(p)=\frac{p}{E_\mu(p)}.
 \label{eq:internalboost}
\end{equation}
Using the variables of Eq.~(\ref{eq:boost}), the accepted range of $u$ at each $x$ is specified by
\begin{equation}
 \begin{split}
 u_-(x,p)&=\max\!\left[-1,\frac{2E_{\min}/(E_\mu x)-1}{\beta}\right],\\
 u_+(x,p)&=\min\!\left[1,\frac{2E_{\max}/(E_\mu x)-1}{\beta}\right].
 \end{split}
 \label{eq:internalulimits}
\end{equation}
The integrands below are set to zero wherever $u_+\leq u_-$. After integrating over the accepted angles we obtain the selected event factor
\begin{equation}
 {\cal N}(p)=\int_0^1 dx\,x^2(3-2x)(u_+-u_-),
 \label{eq:internalN}
\end{equation}
and the amplitude of the energy-dependent modulation
\begin{equation}
 {\cal M}(p)=\frac12\int_0^1 dx\,x^2(2x-1)(u_+^2-u_-^2).
 \label{eq:internalM}
\end{equation}
The factor entering the longitudinal correction follows from Eq.~(\ref{eq:fxu}). Its numerator is
\begin{equation}
 \begin{split}
 {\cal F}(p)=\int_0^1 dx\,x^2(2x-1)\Bigg[&
 \frac{\beta}{2}\left(1-\frac{x}{2}\right)(u_+^2-u_-^2)\\
 &-\frac{x}{6}(u_+^3-u_-^3)\Bigg].
 \end{split}
 \label{eq:internalF}
\end{equation}
Thus the correction and the effective analysing asymmetry are determined by
\begin{equation}
 \langle f_\parallel\rangle_{NA,p}=\frac{{\cal F}(p)}{{\cal M}(p)},\qquad
 A_{\rm eff}(p)=\frac{{\cal M}(p)}{{\cal N}(p)}.
 \label{eq:internalmean}
\end{equation}
The event factors have a common arbitrary normalisation which cancels in their ratios. No transverse contribution is added to the mean correction.

At $300$ MeV/$c$ we recover $\langle f_\parallel\rangle_{NA}=0.20627$. The maximum sensitivity of the internal comparison is found at $p=221.4$ MeV/$c$, corresponding to a total muon energy of $245.3$ MeV. The numerical integrals at these two momenta give
\begin{equation}
 \begin{array}{c|ccc}
 p\ (\mathrm{MeV}/c)&{\cal N}&A_{\rm eff}&\langle f_\parallel\rangle_{NA}\\ \hline
 300.0&0.125086&0.419271&0.206272\\
 221.4&0.022428&0.740309&0.043471
 \end{array}
 \label{eq:internalvalues}
\end{equation}
{\color{black}\hypersetup{citecolor=black}Our $A_{\rm eff}(300)=0.419271$ reproduces the E34 design value $0.42$ \cite{abe2019}, although the detector acceptance is not modelled explicitly.}

The reduced number of selected positrons at the lower momentum is partly compensated by a larger analysing asymmetry. Keeping the correction coefficient of Eq.~(\ref{eq:delta}), the change in the inferred anomaly is
\begin{equation}
 \begin{split}
 a_\mu(221.4)-a_\mu(300)
 &=7.0\times10^{-9}\bigl(0.043471-0.206272\bigr)\\
 &=-1.140\times10^{-9}=-0.977\ \mathrm{ppm}.
 \end{split}
 \label{eq:internaldifference}
\end{equation}
The quoted momenta and integrated quantities refer to the idealised decay distribution with a fixed positron threshold.

To estimate the experimental sensitivity, we assume that each momentum setting receives a complete exposure and has the same number of stored muons before the positron selection. We use the accepted event factor and analysing asymmetry to scale the statistical precision from the reference configuration. The time available for the frequency measurement also scales with the muon lifetime in the laboratory. These assumptions give
\begin{equation}
 \sigma_{\rm stat}(p)=\sigma_{\rm stat}(300)
 \frac{\gamma(300)}{\gamma(p)}
 \frac{A_{\rm eff}(300)}{A_{\rm eff}(p)}
 \sqrt{\frac{{\cal N}(300)}{{\cal N}(p)}}.
 \label{eq:internalsigma}
\end{equation}
We examine the initial E34 design sensitivity of $450$ ppb statistical and the more optimistic value of $100$ ppb statistical \cite{otani2022,e34status}. In both cases we allow a common systematic component of $70$ ppb in each measurement. If it is fully correlated between the two settings, it cancels in their difference. The relevant error and significance are then
\begin{equation}
 \begin{split}
 \sigma_{\Delta}(p)&=\sqrt{\sigma_{\rm stat}^2(300)+\sigma_{\rm stat}^2(p)},\\
 Z(p)&=\frac{|a_\mu(p)-a_\mu(300)|}{a_\mu\,\sigma_{\Delta}(p)\times10^{-6}},
 \end{split}
 \label{eq:internalsignificance}
\end{equation}
where $\sigma_\Delta$ is expressed in ppm. The factor $10^{-6}$ converts ppm to a relative uncertainty. For the uncertainties of the individual measurements shown in the figures, the $70$ ppb systematic contribution is included in quadrature.

At the optimal momentum, Eq.~(\ref{eq:internalsigma}) gives $0.780$ ppm for the initial sensitivity and $0.173$ ppm for the optimistic sensitivity. The results of the comparison are
\begin{equation}
 \begin{array}{c|ccc}
 \sigma_{\rm stat}(300)&\sigma_\Delta&\Delta a_\mu\ (\mathrm{ppm})&Z_{\max}\\ \hline
 450\ \mathrm{ppb}&0.901\ \mathrm{ppm}&-0.977\pm0.901&1.09\\
 100\ \mathrm{ppb}&0.200\ \mathrm{ppm}&-0.977\pm0.200&4.88
 \end{array}
 \label{eq:internaloutcomes}
\end{equation}
{\color{black}If the $70$ ppb systematics at the two momentum settings are uncorrelated, at $p=221.4$ MeV/$c$ we obtain $\sigma_\Delta=\sqrt{0.200^2+2(0.070)^2}=0.223$ ppm and $Z=4.38\sigma$, rather than $4.88\sigma$ for fully correlated systematics.}\par

\begin{figure}[!htbp]
\centering
\includegraphics[width=0.70\textwidth]{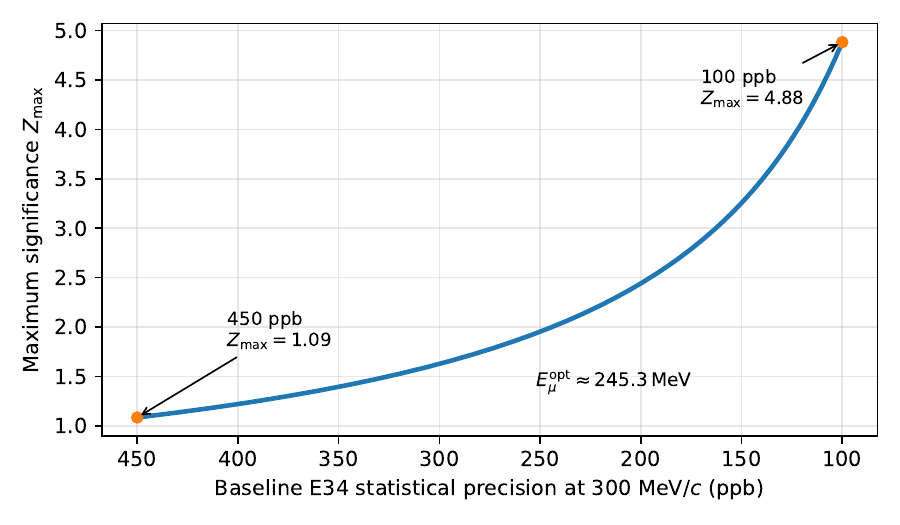}
\caption{Maximum projected significance of the internal two-momentum comparison as a function of the E34 statistical precision at 300 MeV/$c$. The positron energy threshold is fixed at 200 MeV and common systematic uncertainties are assumed to cancel. The optimal muon energy remains approximately 245.3 MeV throughout the interval.}
\label{fig:zmaxprecision}
\end{figure}

\begin{figure}[p]
 \centering
 \includegraphics[width=0.98\textwidth]{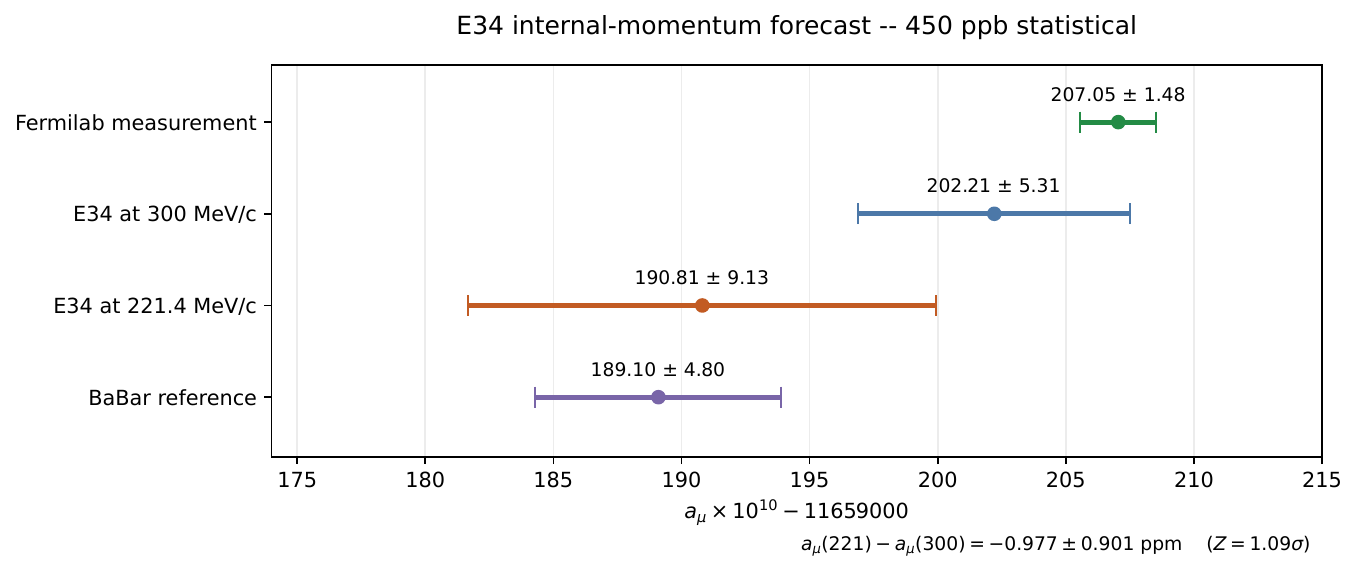}
 \par\medskip
 \includegraphics[width=0.88\textwidth]{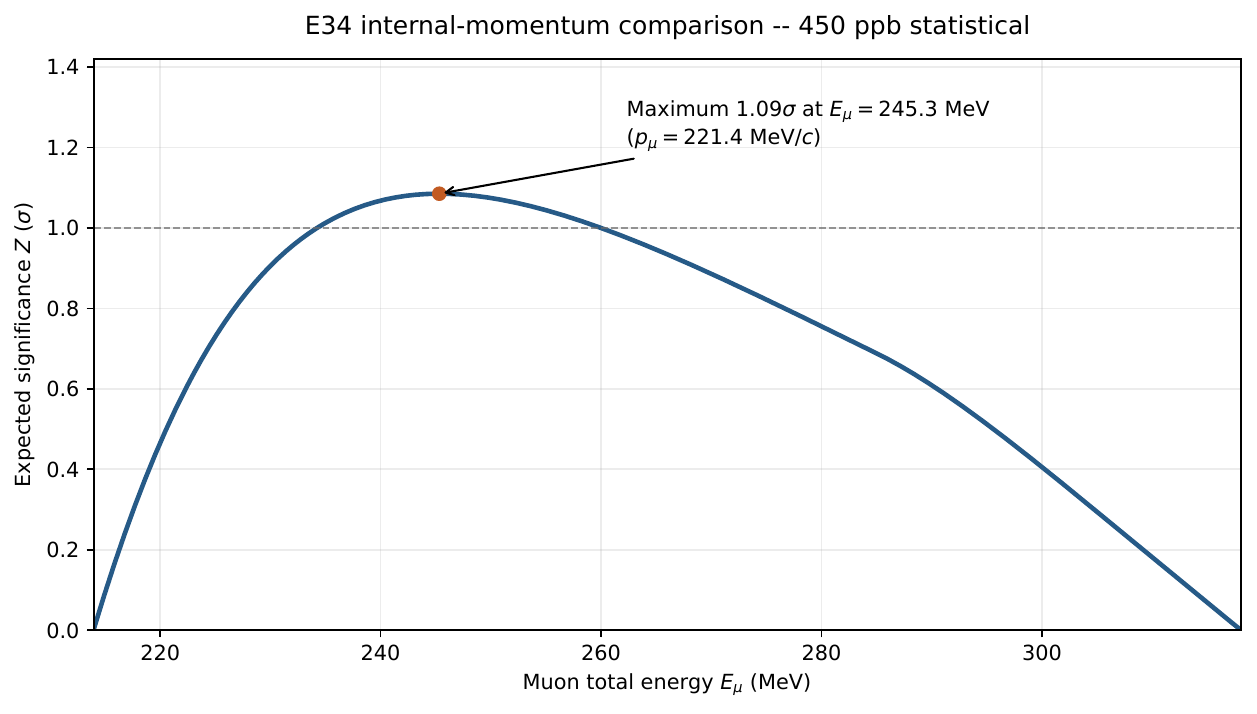}
 \caption{Internal comparison with the initial E34 statistical sensitivity of $450$ ppb at $300$ MeV/$c$. The upper panel shows the two E34 forecasts with their individual one-standard-deviation uncertainties. The Fermilab measurement and the BaBar-based value are included for visual comparison and do not enter the internal difference. The lower panel shows the significance as a function of the total muon beam energy. The $70$ ppb common systematic uncertainty is included in the individual E34 error bars but cancels in the internal comparison.}
 \label{fig:internalinitial}
\end{figure}

\begin{figure}[p]
 \centering
 \includegraphics[width=0.98\textwidth]{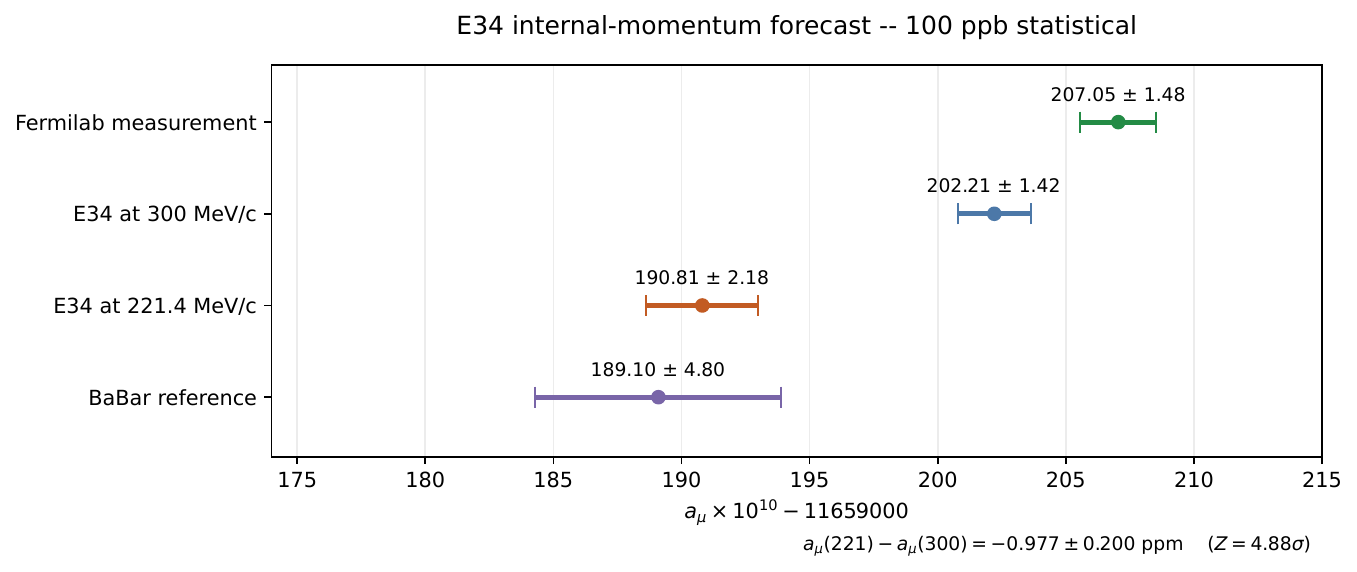}
 \par\medskip
 \includegraphics[width=0.88\textwidth]{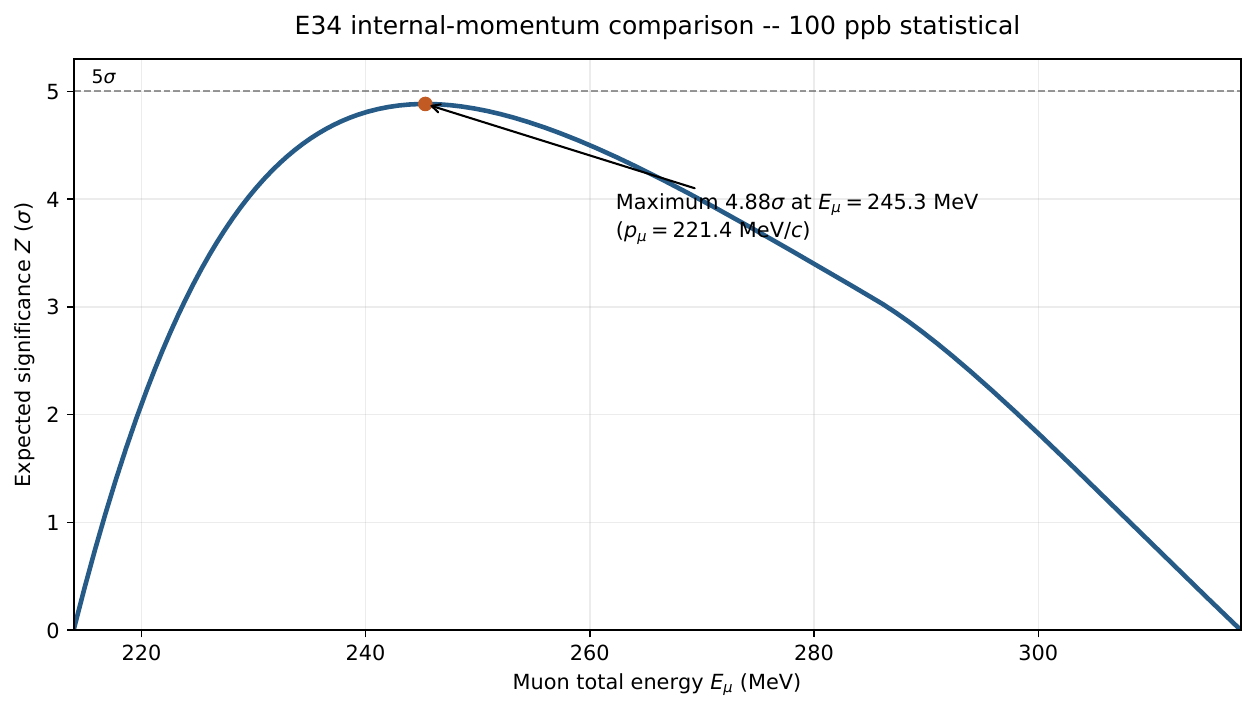}
 \caption{The same comparison for the optimistic E34 statistical sensitivity of $100$ ppb at $300$ MeV/$c$. The longitudinal correction is unchanged, while the reduced statistical uncertainty brings the maximum projected significance to about $5\sigma$ if the systematic contribution cancels. The positions of the Fermilab and BaBar points are displayed only as external references.}
 \label{fig:internaloptimistic}
\end{figure}

The maximisation gives the same lower momentum for both statistical precision scenarios because the common systematic contribution is absent from the difference and the statistical precision changes by an overall factor. Figures~\ref{fig:internalinitial} and \ref{fig:internaloptimistic} show the predicted anomaly measurements together with the dependence of the significance on the beam energy. 
The dependence of the maximum significance on the expected experimental precision is shown in Fig.~\ref{fig:zmaxprecision}.

For the bar plots we use the final Fermilab result as a common reference for the absolute E34 forecast at $300$ MeV/$c$, following the preceding section. On the axis $a_\mu\times10^{10}-11659000$, the resulting central values are $202.21$ and $190.81$ for E34 at $300$ and $221.4$ MeV/$c$, respectively. The individual projected uncertainties are $5.31$ and $9.13$ in the initial scenario and $1.42$ and $2.18$ in the optimistic scenario. The common Fermilab reference uncertainty is excluded from these E34 instrumental bars, as it cancels from the predicted difference between the two E34 configurations.

The comparison would no longer have this sensitivity if the common systematics changed appreciably with the storage orbit. The idealised treatment also assumes that the flux of stored muons, detector response and effective event acceptance can be preserved at the lower momentum. The E34 apparatus has been designed around $300$ MeV/$c$ and the result does not establish that the alternative setting can be achieved with its present injection system. A dedicated apparatus with two muon momenta would offer a more direct test of the energy dependence proposed here.

An alternative is to preserve the nominal storage radius while
reducing the magnetic field according to
$B(p)=3\,\mathrm{T}\,[p/(300\,\mathrm{MeV}/c)]$.
Since $\omega_a\propto B$, the statistical scaling in
Eq.~(4.9) acquires an additional factor $3\,\mathrm{T}/B(p)$.
Repeating the optimisation gives $p=229.1\,\mathrm{MeV}/c$
and $B=2.291\,\mathrm{T}$, with a predicted difference of
$-0.865$ ppm. The maximum significance decreases to
$0.87\sigma$ and $3.91\sigma$ for the initial and optimistic
statistical sensitivities, respectively, assuming fully
correlated systematics. With independent systematic
uncertainties of 70 ppb per measurement, the latter
decreases to $3.59\sigma$. This configuration preserves
the storage geometry but requires a new assessment of
magnetic-field uniformity, injection and systematic
correlations.

\section{Final remarks}
We have shown that the proposed neutrino angular-momentum contribution is sensitive to the decay kinematics. Fermilab's highly relativistic beam makes the forward approximation accurate over the selected positron energies. At J-PARC the muon energy is about ten times smaller and the longitudinal neutrino-pair momentum entering the mean correction decreases.

With the same theoretical weak-decay spectrum and the same $NA$ prescription in both experiments, we find a difference of $-0.415\pm0.176$ ppm in the optimistic E34 scenario. The expected $2.36\sigma$ separation gives a possible signature of the effect through the comparison of two muon beams with substantially different boosts. It remains a model-dependent forecast, particularly because the experimental acceptance and the final E34 analysis weights have not yet been included. The use of Fermilab's published A-method result as the reference is another approximation whose numerical impact is smaller than the present comparison uncertainty.

The advantage of this test is its independence from the hadronic evaluation of the magnetic anomaly. A measured E34 value near the Fermilab result would reduce the room for the proposed boost-dependent correction, while a lower value of the size estimated here would motivate a direct analysis of the energy dependence with the experimental spectra of both collaborations.

A further possibility arises from the freedom to operate the J-PARC experiment at muon momenta below the nominal value of $300$ MeV/c. We have investigated the sensitivity of a comparison between two measurements performed at different momenta, keeping the positron energy threshold fixed at 200 MeV. The competition between the increasing magnitude of the predicted correction and the decreasing statistical precision produces a maximum sensitivity at approximately $221$ MeV/c. At this momentum, we predict a difference of $-0.977 \pm 0.200$ ppm for the most optimistic sensitivity, corresponding to $4.88\sigma$. Although such a configuration would require substantial modifications to the proposed experimental apparatus, the possibility of reaching a significance around $5\sigma$ may provide motivation for future experiments designed specifically to investigate the momentum dependence of the muon magnetic moment measurement.

\section*{Acknowledgements}
FCS thanks FAPESP (S\~ao Paulo, Brazil) for his PhD grant number 2024/19103-9. SC is supported by CNPq (Brazil) with grant number 308518/2023-3.


\begin{thebibliography}{99}

\bibitem{bennett} G. W. Bennett \emph{et al.} (Muon $g-2$ Collaboration), Phys. Rev. D \textbf{73}, 072003 (2006).
\bibitem{aguillard2025} D. P. Aguillard \emph{et al.} (Muon $g-2$ Collaboration), Phys. Rev. Lett. \textbf{135}, 101802 (2025).
\bibitem{fnalreport} Muon $g-2$ Collaboration, \emph{Final Report on the Measurement of the Positive Muon Anomalous Magnetic Moment at Fermilab to 127 ppb}, arXiv:2606.17323 (2026).
\bibitem{aoyama2020} T. Aoyama \emph{et al.}, Phys. Rep. \textbf{887}, 1 (2020).
\bibitem{davier2020} M. Davier, A. Hoecker, B. Malaescu and Z. Zhang, Eur. Phys. J. C \textbf{80}, 241 (2020).
\bibitem{babar} B. Aubert \emph{et al.} (BaBar Collaboration), Phys. Rev. Lett. \textbf{103}, 231801 (2009).
\bibitem{kloe} A. Anastasi \emph{et al.} (KLOE-2 Collaboration), JHEP \textbf{03}, 173 (2018).
\bibitem{cmd3} F. V. Ignatov \emph{et al.} (CMD-3 Collaboration), Phys. Rev. Lett. \textbf{132}, 231903 (2024).
\bibitem{borsanyi2021} S. Borsanyi \emph{et al.}, Nature \textbf{593}, 51 (2021).
\bibitem{aliberti2025} R. Aliberti \emph{et al.}, Phys. Rep. \textbf{1143}, 1 (2025).
\bibitem{carneiro2023} S. Carneiro and F. C. Sobrinho, Braz. J. Phys. \textbf{53}, 103 (2023).
\bibitem{carneiro2026} S. Carneiro and F. C. Sobrinho, Braz. J. Phys. \textbf{56}, 197 (2026).
\bibitem{carneiro2026pre} S. Carneiro and F. C. Sobrinho, \emph{On trace invariance and energy-dependent entanglement in muon decay}, arXiv:2608.19287 (2026).
\bibitem{abe2019} {\color{black}M. Abe \emph{et al.}, Prog. Theor. Exp. Phys. \textbf{2019}, 053C02 (2019).}
\bibitem{otani2022} M. Otani, Prog. Theor. Exp. Phys. \textbf{2022}, 052C01 (2022).
\bibitem{e34status}
L. Cotrozzi (J-PARC Muon $g-2$/EDM Collaboration),
\emph{Status of the Muon $g-2$/EDM Experiment at J-PARC},
XV International Conference on New Frontiers in Physics
(ICNFP 2026), CERN Indico contribution 7103098.
\bibitem{Landau 4} V. B. Berestetskii, E. M. Lifshitz and L. P. Pitaevskii, {\it Quantum Electrodynamics} (Pergamon, 1982), Sec.~1.
\bibitem{landau1975} L. D. Landau and E. M. Lifshitz, \emph{The Classical Theory of Fields} (Pergamon, 1975), Sec. 14.
\bibitem{giunti2015} C. Giunti and A. Studenikin, Rev. Mod. Phys. \textbf{87}, 531 (2015).
\bibitem{tiomno1949} J. Tiomno and J. A. Wheeler, Rev. Mod. Phys. \textbf{21}, 144 (1949).
\bibitem{michel1950} L. Michel, Proc. Phys. Soc. A \textbf{63}, 514 (1950).
\end{thebibliography}
\end{document}